\documentclass[a4paper,12pt]{article}
\usepackage{graphicx}
\usepackage{amssymb,amsmath,latexsym}
\usepackage[left=2cm,right=2cm, top=2cm,bottom=2cm,bindingoffset=0cm]{geometry}
\usepackage[T2A]{fontenc}
\usepackage{graphicx}
\usepackage{cite}
\usepackage{amssymb,amsmath,latexsym}%,enumerate}
\usepackage{caption}
\usepackage{authblk}
\usepackage{color}

\usepackage{epsfig}

\begin{document}

\title{Bubbling transition as a mechanism of destruction of synchronous oscillations of identical microbubble contrast agents}% Force line breaks with \\

\author[1]{Ivan R. Garashchuk}
\author[1]{Dmitry I. Sinelshchikov}
\affil[1]{National Research University Higher School of Economics, Moscow, Russia}

\maketitle

\begin{abstract}
We study the process of destruction of synchronous oscillations in a model of two interacting microbubble contrast agents exposed to an external ultrasound field. Completely synchronous oscillations in this model are possible in case of identical bubbles, when the governing system of equations possess a symmetry leading to the existence of a synchronization manifold. Such synchronous oscillations can be destructed without breaking the corresponding symmetry of the governing dynamical system. Here we describe the phenomenological mechanism responsible for such destruction of synchronization and demonstrate its implementation in the studied model.
We show that the appearance and expansion of transversally unstable areas in the synchronization manifold leads to transformation of a synchronous chaotic attractor into a hyperchaotic one. We also demonstrate that this bifurcation sequence is stable with respect to symmetry breaking perturbations.
\end{abstract}

\section{Introduction}

Here we consider a model describing oscillations of two interacting microbubble contrast agents. Microbubble contrast agents are micrometer size encapsulated gas bubbles which are used for enhancing ultrasound visualisation \cite{Szabo,Goldberg,Hoff}. There are also other applications of such bubbles such as targeted drug delivery and noninvasive therapy \cite{Klibanov2006,Coussios2008}.

It is known that the type of dynamics of contrast agents may be important for particular applications (see, e.g. \cite{Hoff,Carroll2013}). For instance, chaotic dynamics may be beneficial for ultrasound visualization, while regular dynamic may be required for targeted drug delivery. Moreover, large amplitude hyperchaotic oscillations may be utilized for the purpose of quick destruction of the bubbles, e.g. in some drug delivery related applications\cite{Klibanov2006}. Therefore, it is necessary to determine both possible type of bubbles oscillations and their bifurcations. On the other hand, in a number of works it was shown that the contrast agents dynamics may be complicated and strongly depend on both control parameters and initial conditions (see, e.g.\cite{Parlitz1990,Behina2009,Carroll2013,Garashchuk2018,Macdonald2006,Garashchuk2019,Garashchuk2020}). In works\cite{Parlitz1990,Behina2009,Carroll2013,Garashchuk2018} the case of one microbubble was studied and it was demonstrated that the dynamics may be either regular or chaotic and multistable in both cases, while, e.g. in \cite{Macdonald2006,Garashchuk2019,Garashchuk2020}, interactions between bubbles were taken into account.

In particular, in \cite{Garashchuk2019} it was found that the dynamics of two interacting identical contrast agents may be regular, quasiperiodic, chaotic and hyperchaotic. Furthermore, authors of  Ref. \cite{Garashchuk2019} demonstrated that oscillations of bubbles may be both synchronous and asynchronous. It is worth noting that a synchronization manifold exist only in the case of two identical bubble, when the governing system of equations possesses a symmetry\cite{Garashchuk2019,Garashchuk2020}. Thus, complete synchronization is possible only in  the symmetrical case. In\cite{Garashchuk2020} authors studied how symmetry breaking affects various synchronous and asynchronous attractors. However, the destructions of synchronous oscillations without symmetry breaking in the model of two interacting contrast agents has not been studied previously.

Therefore, the main aim of this work is to study the process of destruction of synchronous oscillations in a system of two identical interacting contrast agents.  We demonstrate that the typical mechanism responsible for this is the bubbling transition \cite{Ashwin94,Ott1994,Ashwin96,Ott1996,Pikovsky,Ott2003} that was observed in various dynamical systems\cite{Kurths2005,Saha17}. The key point in this scenario is the appearance of transversally unstable saddle orbits inside the synchronization manifold leading to the emergence of transversally unstable areas there and, eventually, to the disappearance of the synchronous attractor. At first, a trajectory can escape the synchronization manifold for a short time. Then, the more transversally unstable orbits emerge in the synchronization manifold, the more frequent and long these escapes become. Finally, the trajectory spends the majority of time outside the synchronization manifold and the attractor becomes hyperchaotic. Since this scenario leads to a hyperchaotic attractor, the bubbling transition can be considered as another route to hyperchaos that is alternative to the scenario involving the emergence of the discrete Shilnikov attractor\cite{Garashchuk2019} (see also\cite{Stankevich2018,Stankevich2019,Kazakov2020,Kazakov2021}).

It is worth noting that in this work we consider complete synchronization of two chaotic systems (see, e.g. \cite{Pikovsky84,Afraimovich86,Carroll90,Pikovsky96,Carroll97,Pikovsky}), each of which represents forced oscillations of a microbubble contrast agent. Single forced model without coupling would demonstrate chaotic behaviour in the considered regions of the control parameters.

The rest of this work is organized as follows. In the next section we introduce the governing dynamical system and discuss some of its properties. In Section \ref{sec:scenario} we propose a phenomenological scenario of the destruction of synchronous oscillations in the symmetrical case of considered model. Section \ref{sec:implement} is devoted to numerical demonstration of the implementation of the proposed scenario. In the last Section we briefly summarize and discuss our results.

\section{Main system of equations}
We consider the following system of equations \cite{Takahira1995,Harkin2001,Ida2002,Doinikov2004,Dzaharudin2013,Macdonald2006}

\begin{eqnarray}
\label{eq:eq1}
    \left(1-\frac{\dot{R_{1}}}{c}\right)R_{1} \ddot{R_{1}} +\frac{3}{2}\left(1-\frac{\dot{R_{1}}}{3c}\right)\dot{R}_{1}^{2}=\frac{1}{\rho}\left[1+\frac{\dot{R}_{1}}{c}+\frac{R_{1}}{c}\frac{d}{dt}\right]P_{1}-\frac{d}{dt}\left(\frac{R_{2}^{2}\dot{R}_{2}}{d}\right), \vspace{0.1cm} \nonumber \\ %\frac{d (R_{2}^{2}\dot{R}_{2})}{dt} \frac{R_{2}^{2}\ddot{R}_{2}+2 R_{2} \dot{R}_{2}^{2}}{d}
     \left(1-\frac{\dot{R_{2}}}{c}\right)R_{2} \ddot{R_{2}} +\frac{3}{2}\left(1-\frac{\dot{R_{2}}}{3c}\right)\dot{R}_{2}^{2}=\frac{1}{\rho}\left[1+\frac{\dot{R}_{2}}{c}+\frac{R_{2}}{c}\frac{d}{dt}\right]P_{2}
     -\frac{d}{dt}\left(\frac{R_{1}^{2}\dot{R}_{1}}{d}\right), % \\% -\frac{R_{1}^{2}\ddot{R}_{1}+2 R_{1} \dot{R}_{1}^{2}}{d},
\end{eqnarray}
where
\[
P_{i}=\left(P_{0}+\frac{2\sigma}{R_{i0}}\right)\left(\frac{R_{i0}}{R_{i}}\right)^{3\gamma}-\frac{4\eta_L \dot{R}_{i}}{R_{i}}-\frac{2\sigma}{R_{i}}-P_{0}-4\chi\left(\frac{1}{R_{i0}}-\frac{1}{R_{i}}\right)-4\kappa_{S}\frac{\dot{R}_{i}}{R_{i}^{2}}-P_{ac}\sin(\omega t), \quad i=1,2.
\]

System \eqref{eq:eq1} describes oscillations of two encapsulated gas bubbles, which interact through the pressure field, in a compressible viscous liquid. Here $R_{1}$ and $R_{2}$ are radii of the bubbles, $d$ is the distance between the centers of bubbles, $P_{\mathrm{stat}} = 100$ kPa is the static pressure, $P_v = 2.33$ kPa is the vapor pressure, $P_0 = P_{\mathrm{stat}} - P_v$, $P_{ac}$ is the magnitude of the pressure of the external field, $\sigma = 0.0725$ N/m is the surface tension, $\rho= 1000$ kg/m$^3$ is the density of the liquid, $\eta_{L} = 0.001$ Ns/m$^3$ is the viscosity of the liquid, $c = 1500$ m/s is the sound speed, $\gamma = 4/3$ is the polytropic exponent (the process is assumed to be adiabatic), $\chi$ and $\kappa_{s}$ denote the shell elasticity and shell surface viscosity respectively. Dynamical system \eqref{eq:eq1} takes into account liquid's compressibility according to the Keller--Miksis\cite{Keller1980} model and bubbles' shell according to the de-Jong model\cite{deJong1992, Marmottant2005, Doinikov2013}.

We use the following values of the parameters $\chi = 0.22$ N/m and $\kappa_S = 2.5\cdot 10^{-9}$ kg/s that correspond to the SonoVue contrast agent with equilibrium radius $R_{i0}=1.72 \mu$m \cite{Tu2009}.

In this work we study the case of bubbles with the same equilibrium radii, i.e. $R_{10}=R_{20}=R_{0}$. Under this condition the system \eqref{eq:eq1} is symmetrical with respect to transformation $R_1 \longleftrightarrow R_2$. This leads to the existence of the synchronization manifold, defined by equations $R_1 = R_2, \dot{R}_1 = \dot{R}_2$, i.e. it is a three-dimensional plane in a five-dimensional phase space. Below we show that a synchronous attractor lying inside this manifold can be destructed through the bubbling transition mechanism. We also demonstrate that our results are stable with respect to the symmetry breaking perturbation.

Parameters $P_{ac}$ and $\omega$ can be naturally considered as control ones, since they define properties of the external ultrasound field. The parameter $d$ is defined by the density of the injected bubbles cluster and is also considered as control one. In this work, we focus on the influence of the parameter $d$ on the dynamics and fix $P_{ac}$ and $\omega$ as follows $P_{ac} = 1.6$ MPa and $\omega = 2.87 10^7$ s$^{-1}$. These values of $P_{ac}$ and $\omega$ correspond to biomedically relevant regions of these parameters \cite{Hoff}.

We perform all calculations in the following non--dimensional variables $R_{i}=R_{10}r_{i}$, $t=\omega_{0}^{-1}\tau$, where $\omega_{0}^{2}=3\kappa P_{0}/(\rho R_{10}^2)+2(3\kappa-1)\sigma/R_{10}+4\chi/R_{10}$ is the natural frequency of bubble oscillations. We also denote the non-dimensional radial speeds of the bubbles' surfaces by $u_{1}$ and $u_{2}$.  We use the fourth-fifth order Runge--Kutta method \cite{Cash1990} for finding numerical solutions of the Cauchy problem for \eqref{eq:eq1}. For the calculations of the Lyapunov spectra we use the standard algorithm by Bennetin et al \cite{Benettin1980}, modified according to the specifics of our problem, see Sec. \ref{sec:implement} for details. Poincar\'e maps are constructed as a cross-section at every period of the external pressure field, i.e. we use stroboscopic section.

\section{Bubbling transition scenario}
\label{sec:scenario}

In this section we propose a phenomenological scenario of the destruction of a synchronous attractor in model \eqref{eq:eq1}. This scenario is based on the bubbling transition mechanism\cite{Ashwin94,Ott1994,Ashwin96,Ott1996,Pikovsky}.

\begin{figure}[!ht]
\center{\includegraphics[width=0.5\textwidth]{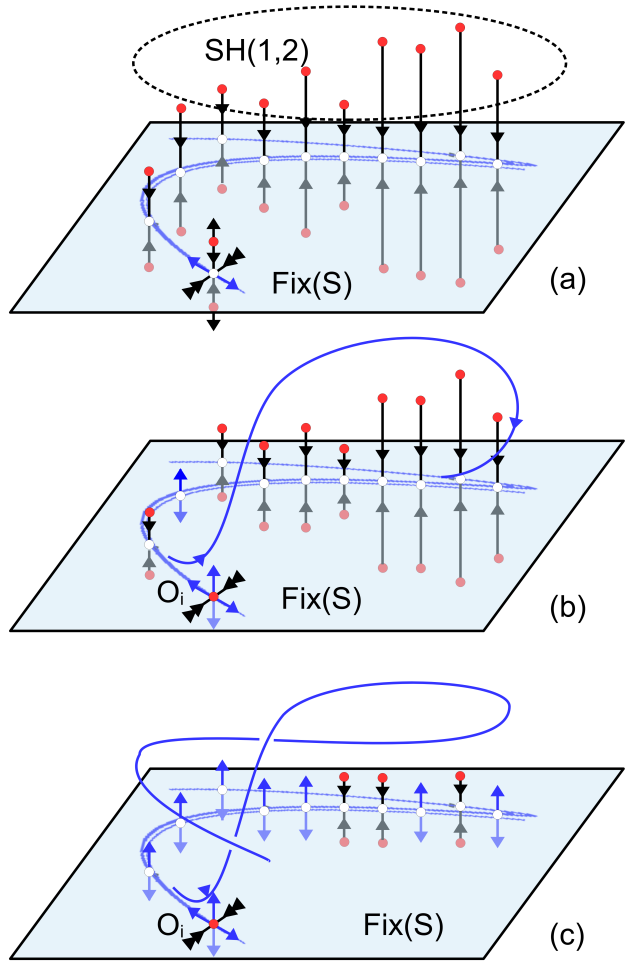}
}
\caption{Scheme of the main steps of the bubbling transition scenario. (a) Strongly synchronous state, (b) weakly synchronous state, (c) weakly asynchronous state.} % Note that on this scheme we use representation on Poincar\'e map is used instead of the phase space.}
\label{fig0}
\end{figure}

We consider a synchronous chaotic attractor in system \eqref{eq:eq1}, belonging to a set $Fix(S)$, obtained via a cascade of period-doubling bifurcations of a synchronous limit cycle. Suppose also we have a non-attractive hyperchaotic hyperbolic set $SH(2,2)$ near $Fix(S)$, see Fig. \ref{fig0}a. Notice that for better clarity we demonstrate the Poincar\'e map instead of the full phase space. Thus, for instance, points in Fig. \ref{fig0} correspond to periodic trajectories of the original system. Let us also remark that it does not matter, how the set $SH(2,2)$ is created (e.g. it is initially non-attractive, or appears after a crisis of a hyperchaotic attractor).

There is an infinite number of saddle limit cycles of $(3,1)$-type inside the synchronous chaotic attractor, which are transversally stable and have one-dimensional unstable manifold embedded inside the synchronization manifold (see white points on $Fix(S)$ in Fig. \ref{fig0}a). In this case we observe strong synchronization\cite{Pikovsky}, since all periodic orbits are transversally stable. On the other hand, set $SH(2,2)$ outside of the synchronization manifold contains infinitely many periodic saddle orbits of $(2,2)$-type (see red points in Fig. \ref{fig0}a). Due to the symmetry of system \eqref{eq:eq1} ($R_1 \longleftrightarrow R_2$), set $SH(2,2)$ consists of two symmetrical parts above and below $Fix(S)$. The set $SH(2,2)$ is hyperchaotic because two-dimensional unstable manifolds of $(2,2)$ saddle periodic orbits result in two directions of hyperbolic instability when a trajectory passes close to them.  This will be numerically demonstrated in Section \ref{sec:implement}.

Then, if we increase a control parameter, set $SH(2,2)$ can approach $Fix(S)$ and unstable manifolds of its $(2, 2)$-type saddle orbits can form heteroclinic manifolds with transversally stable $(3, 1)$-type orbits lying inside $Fix(S)$ and forming the skeleton of the synchronous attractor. Eventually, at a certain value of the control parameter these transversally stable $(3, 1)$ saddle orbits from $Fix(S)$ collide with $(2, 2)$-type saddles from $SH(2,2)$ merging via subcritical pitchfork bifurcations and, thus, creating transversally unstable $(2, 2)$-type orbits $O_i$ inside $Fix(S)$ (see Fig. \ref{fig0}b). One unstable invariant manifold of $O_i$ is embedded inside $Fix(S)$, while another one is transversal to it. Since the most of the orbits belonging to $Fix(S)$ are still transversally stable, the ejection of a trajectory from $Fix(S)$ is a rare event, which can be considered as an extreme event \cite{Saha17}. Moreover, after escaping from $Fix(S)$ through the transversally unstable invariant manifold of $O_i$, the trajectory goes through $SH(2,2)$ and returns back to $Fix(S)$ through a transversally stable area of $Fix(S)$. At this point the synchronous attractor becomes a Milnor attractor with riddled basin\cite{Pikovsky}. Below, by computing transversal Lyapunov exponents\cite{Carroll97,Pikovsky}, we demonstrate that it is stable on average. Therefore, we have a weakly synchronous state at these values of the control parameter.

Consequently, an orbit naturally splits into two phases: synchronous and asynchronous. There are two logically distinct sets inside the attractive area of the phase space, through which a trajectory goes: chaotic set inside $Fix(S)$ and hyperchaotic saddle set $SH(2,2)$, and attractor as a whole consists of synchronous and asynchronous components. If we further increase the value of the control parameter, the subcritical pitchfork bifurcations continue to happen, and more and more transversally unstable synchronous orbits of $(2, 2)$-type appear in $Fix(S)$. As the joint contribution of such transversally unstable orbits becomes more significant, substantial areas of $Fix(S)$ become transversally unstable. Therefore, a trajectory spends more and more time outside the synchronization manifold. At a certain point a balance can be reached: trajectory spends roughly equal time inside and outside $Fix(S)$, and both components have similar impact on the behavior of a trajectory as the whole. At the point when the largest transversal Lyapunov exponent becomes positive, the synchronous set $Fix(S)$ turns transversally unstable on average, while there still exist transversally stable orbits inside the synchronization manifold. Here we observe a weakly asynchronous state.
Since the asynchronous component is hyperchaotic, this weakly asynchronous state can also be hyperchaotic. The implementation of this scenario in system \eqref{eq:eq1} will be demonstrated numerically in Sec. \ref{sec:implement}.

Increasing the control parameter further could lead to the blow-out bifurcation \cite{Ott1994,Ott1996,Saha17}: almost all of the orbits that used to form the skeleton of the synchronous attractor become transversally unstable. The manifold $Fix(S)$ becomes completely transversally unstable, and a trajectory never comes back to it. The set $SH(2, 2)$ becomes attractive, and the attractor becomes fully asynchronous and hyperchaotic. However, in the case of system \eqref{eq:eq1} along the one-parametric route studied in this paper, the blow out bifurcation does not happen since the hyperchaoic attractor disappears earlier due to a crisis.

\section{Implementation of the bubbling transition scenario}
\label{sec:implement}

\begin{figure}[!ht]
\center{
\includegraphics[width=0.5\textwidth]{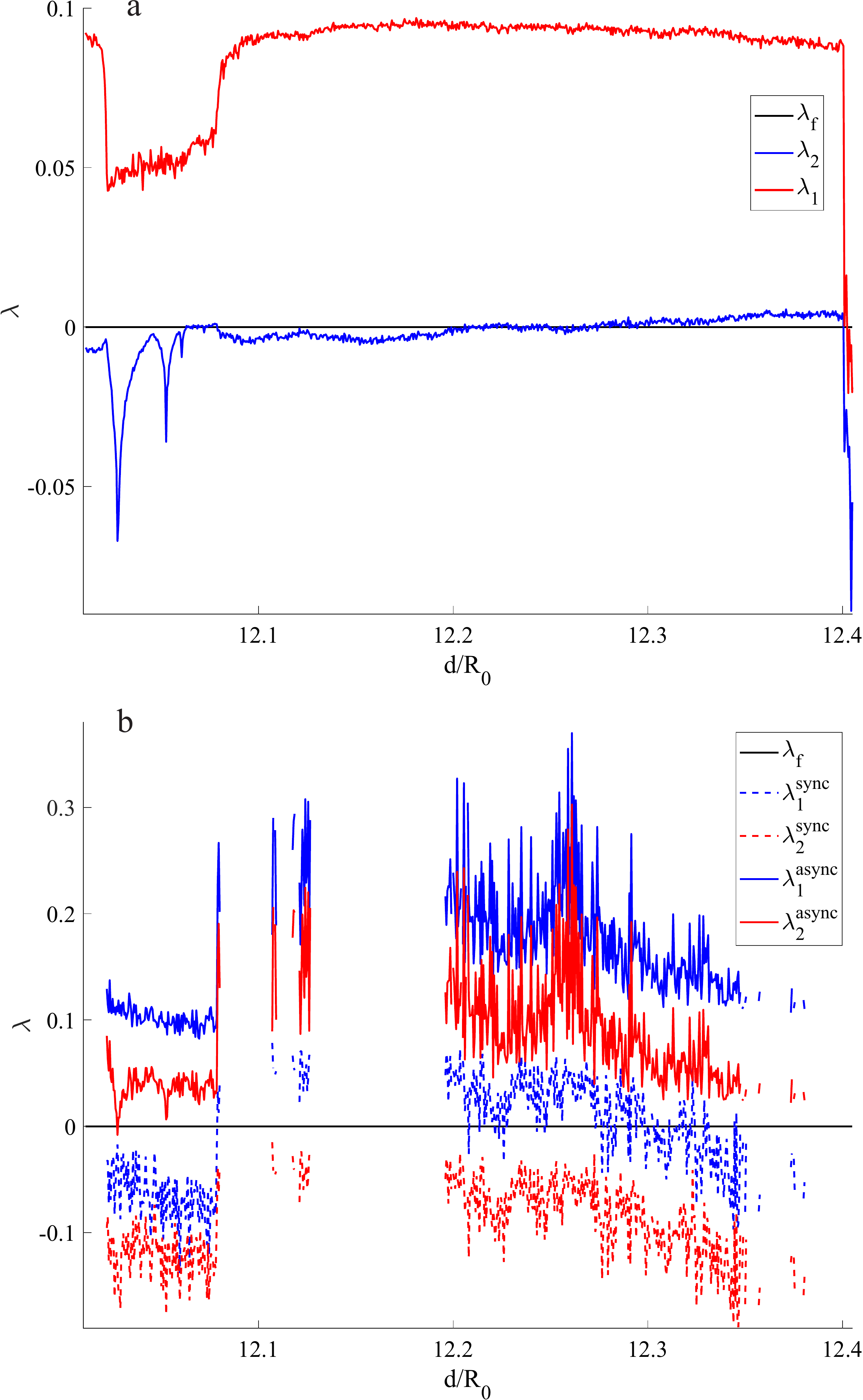}
}
\caption{The dependence of the three largest Lyapunov exponents on $d/R_{10}, \in [12.01, 12.4]$: (a) calculated along the full trajectories, (b) calculated separately for synchronous and asynchronous phases of the trajectories (plotted only for intervals with enough points in both phases).}
\label{fig1a}
\end{figure}

\begin{figure}[!ht]
\center{
\includegraphics[width=0.5\textwidth]{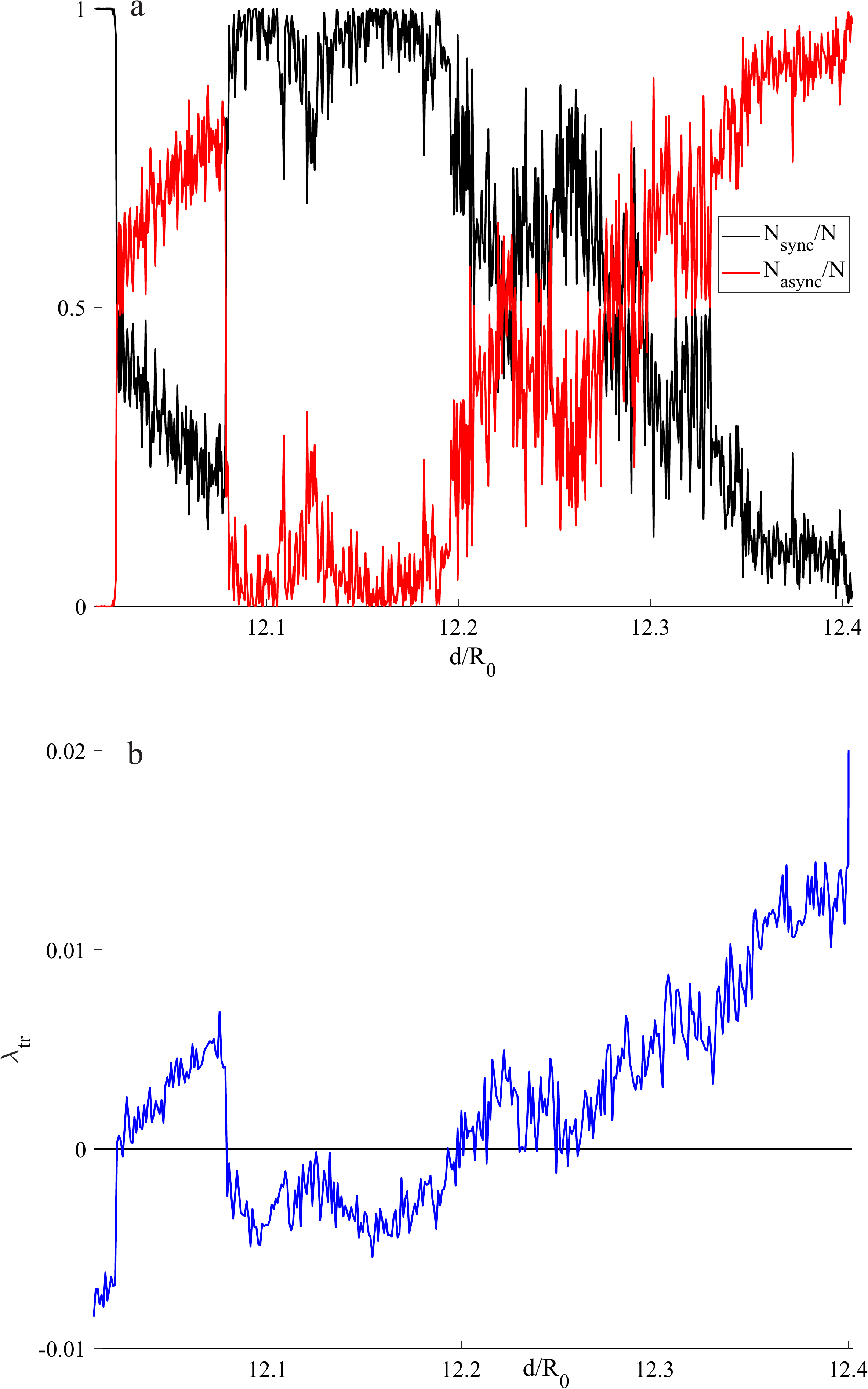}
}
\caption{The dependence of (a) relative numbers of points in synchronous and asynchronous components $N_{async}/N$ and $N_{sync}/N$ on $d/R_{10}$ and (b) the maximal transversal Lyapunov exponent on $d/R_{10}$.}

\label{fig1c}
\end{figure}

In this section we demonstrate the implementation of the above proposed scenario of destruction of a synchronous attractor in model \eqref{eq:eq1}. As we show below, this bifurcations sequence leads to a transformation from a synchronous periodic attractor to a hyperchaotic one, consisting of synchronous and asynchronous parts, in accordance with the hypothesis in Sec. \ref{sec:scenario}. In numerical calculations we split the points of a trajectory into synchronous and asynchronous ones according to the following criterion:  if $|r_1 - r_2| < 10^{-6}, |u_1 - u_2| < 10^{-6}$, then the corresponding point belongs to the synchronization manifold, otherwise it belongs to the asynchronous part of a trajectory. We also modify the scheme of calculating the Lyapunov spectrum in the same way: exponents for synchronous and asynchronous components are calculated separately. Lyapunov exponents of an attractor as a whole can be calculated by averaging along the whole trajectory or as a weighted sum of both components with weights $N_{sync}/N$ and $N_{async}/N$ respectively, where $N$ is total number of points used in calculations and, $N_{sync}$ and $N_{async}$ are numbers of points belonging to synchronous and asynchronous parts of a trajectory, respectively.

We perform calculations of the  Lyapunov spectrum with the following time step $T_l = 0.1 T$ between orthonormalizations, where $T$ is the non-dimensional period of the external pressure field. For each value of the control parameter we use $1.3 \cdot 10^6$ steps of length $T_l$, which corresponds to the non-dimensional time $1.3 \cdot 10^5 T$. We consider our estimations of Lyapunov exponents  statistically correct only if at least $2 \cdot 10^5$ steps belong to each component of the trajectory (or at least $15\%$ of points belong to each phase).  Thus, synchronous and asynchronous parts of the Lyapunov spectrum are demonstrated only for intervals of the control parameter for which the number of points belonging to each component exceeds this threshold.

As a quantitative estimation of transversal stability of the synchronization manifold we use the largest transversal Lyapunov exponent\cite{Pikovsky,Carroll90,Carroll97}. It shows whether an attractor is transversally stable or unstable on average. If this exponent is negative, then the corresponding attractor is transversally stable on average, and otherwise if it is positive. We do not provide the values of the second transversal Lyapunov exponent, because it is always negative with a comparatively large absolute value.
With the help of the combination of the calculated quantities (number of synchronous/asynchronous points, Lyapunov and transversal Lyapunov exponents) we can separate different qualitative types of behavior along the route from synchronous to asynchronous state.  If there are no asynchronous points along a trajectory and the largest transversal Lyapunov exponent is negative, we observe strongly synchronous state.
Suppose that the asynchronous phase of the trajectory is present, but the largest transversal Lyapunov exponent is still negative. Consequently, the synchronous attractor is still transversally stable on average and we have a weakly synchronous state. If there are asynchronous and synchronous phases of a trajectory and the largest transversal Lyapunov exponent is positive, we deal with a weakly asynchronous state. If all the orbits inside the synchronization manifold became unstable, one could observe strongly asynchronous state. However, in the considered one-parametric route the hyperchaoic attractor disappears earlier due to a crisis.

In what follows we fix $\omega$ and $P_{ac}$ and consider $d/R_{10}$ as the control parameter and provide one-parametric maps of quantities characterizing the dynamics of the system  with respect to this parameter.  In Fig. \ref{fig1a}a we present the dependence of Lyapunov exponents calculated along the whole trajectory on the control parameter. To show possible types of behavior in different phases we demonstrate Lyapunov exponents that are calculated separately for synchronous and asynchronous components of this trajectory in Fig. \ref{fig1a}b. We also show how ratios $N_{async}/N$ and $N_{sync}/N$ change with the parameter $d/R_{10}$ in Fig. \ref{fig1c}a. They approximately represent fractions of time that a trajectory spends inside synchronization manifold and outside of it, respectively. The graph of the largest transversal Lyapunov exponent is presented in Fig. \ref{fig1c}b. Once can clearly notice strong correlation between graphs in Fig.-s \ref{fig1c}a,b.

In order to calculate the Lyapunov spectrum for \eqref{eq:eq1} we convert it into five-dimensional autonomous dynamical system. Any attractor of such system, which is not a stationary point, has a zero Lyapunov exponent, that corresponds to the translations along this attractor. We denote this Lyapunov exponent by $\lambda_f$ and plot it on the graphs as a reference to conveniently separate positive and negative values of the exponents. Apart from $\lambda_f$, we present only the two largest exponents, because they help to distinguish between different types of attractors. On the other hand, the other two exponents are always negative (since system \eqref{eq:eq1} is dissipative) and their values do not carry any substantial information. If we presented these Lyapunov exponents in the same plots with the two largest ones, we would significantly decrease the clarity due to the following reason. It would be necessary to change the scale of the graphs greatly in order to fit those large negative values. As a consequence, it would be very difficult to distinguishing between various dynamical regimes since the two largest Lyapunov exponents are often close to zero. To avoid ambiguity, below, we exclude $\lambda_f$ from the consideration, when we discuss numbers of positive and negative exponents. In Tables \ref{tbl0}, \ref{tbl1} we present the values of Lyapunov exponents, the proportion of asynchronous points and the largest transversal Lyapunov exponent for several particular values of parameters, that correspond to certain steps in the phenomenological scenario proposed in the previous section. In these Tables we also exclude $\lambda_f$ from consideration.

%the two largest Lyapunov exponents are often close to zero, and distinguishing between various dynamical regimes while reading the graphs would become much harder.}

One can see that the increase in the parameter $d/R_{10}$ leads to the emerging of asynchronous points even if one starts with synchronous initial conditions. This means that transversally unstable saddle orbits indeed start appearing inside the synchronization manifold and a trajectory may leave the synchronization manifold when it passes close to one of these saddle orbits. Therefore, the suggested splitting of a trajectory on synchronous and asynchronous parts is quite natural. Notice also that the asynchronous phase is virtually always characterized by two positive Lyapunov exponents, which means that the dynamics in the asynchronous phase is hyperchaotic, i.e. two directions of hyperbolic instability are present outside of the synchronization manifold. This supports our hypothesis of the existence of a saddle set outside the synchronization manifold, containing saddle orbits of $(2,2)$-type.

\begin{figure}[!ht]
\center{\includegraphics[width=0.5\textwidth]{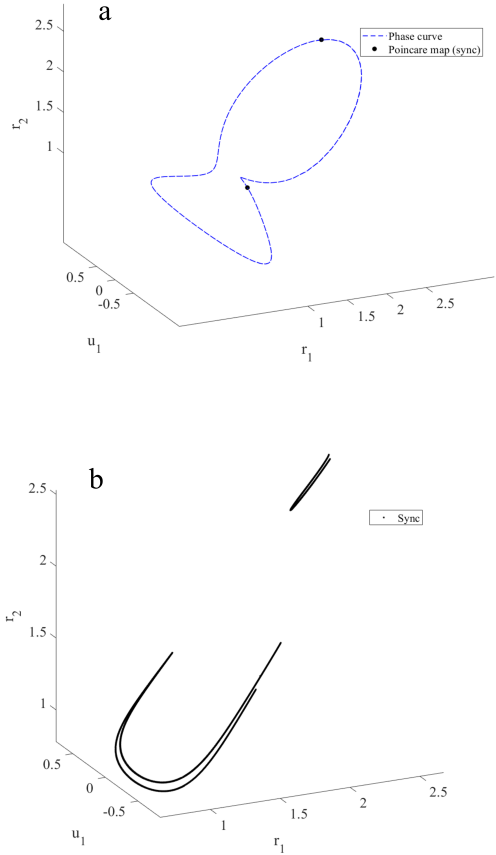}
}
\caption{Projections on the $(r_1, u_1, r_2)$ space of (a) phase curve and Poincar\'e map of the synchronous 2-periodic limit cycle at $d/R_{10} = 7.6$, (b) synchronous chaotic attractor at $d/R_{10} = 12.01$, emerging as a result of a cascade of period-doubling bifurcations of the limit cycle in panel (a) }
\label{fig2}
\end{figure}

\begin{figure}[!ht]
\center{\includegraphics[width=0.5\textwidth]{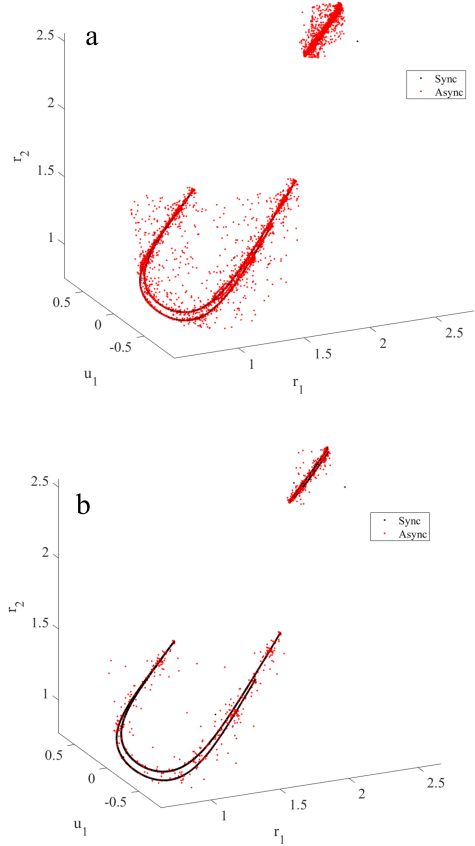}
}
\caption{Projections of Poincar\'e maps of attractors on the $(r_1, u_1, r_2)$ space at (a) $d/R_{10} = 12.05$, $N_{async}/N = 0.7164$ high presence of asynchronous component, (b) $d/R_{10} = 12.12$, $N_{async}/N = 0.2055$ substantial drop in the amount of points in the asynchronous component }
\label{fig3}
\end{figure}

\begin{figure}[!ht]
\center{\includegraphics[width=0.5\textwidth]{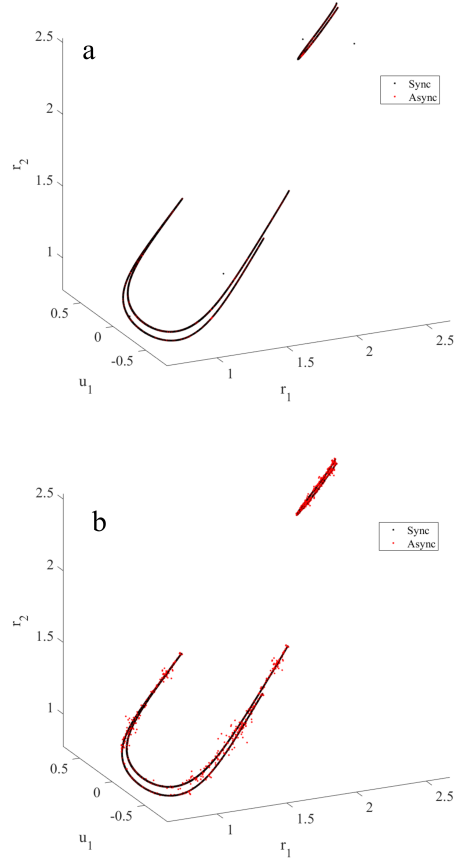}
}
\caption{Projections of Poincar\'e maps of attractors on the $(r_1, u_1, r_2)$ space at (a) $d/R_{10} = 12.18$, $N_{async}/N = 0.037$ the asynchronous component is almost absent and barely visible, (b) $d/R_{10} = 12.20$, $N_{async}/N = 0.150$ asynchronous component becomes a little more noticable}
\label{fig4}
\end{figure}

\begin{figure}[!ht]
\center{\includegraphics[width=0.5\textwidth]{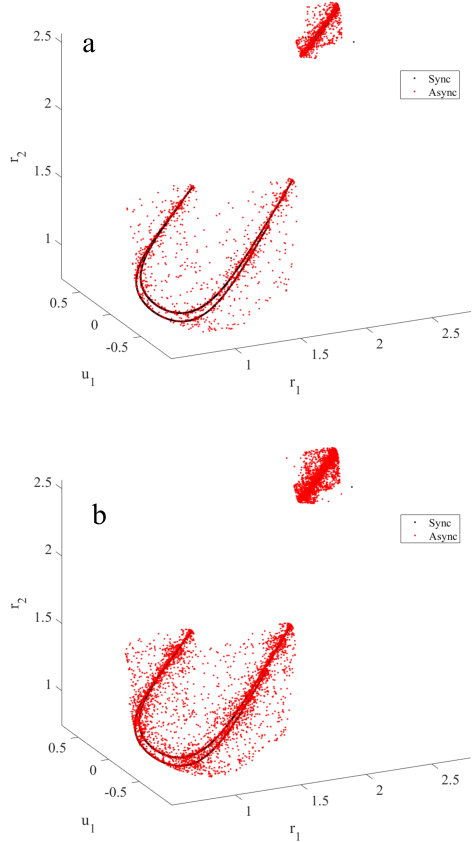}
}
\caption{Projections of Poincar\'e maps of attractors on the $(r_1, u_1, r_2)$ space at (a) $d/R_{10} = 12.27$, $N_{async}/N = 0.338$ there is substantial presence of the asynchronous component, the asynchronous points occur a little bit further away from the synchronization manifold, (b) $d/R_{10} = 12.276$, $N_{async}/N =0.439$, asynchronous points can go further away from the synchronization manifold, and their density has increased}
\label{fig5}
\end{figure}

\begin{figure}[!ht]
\center{\includegraphics[width=0.5\textwidth]{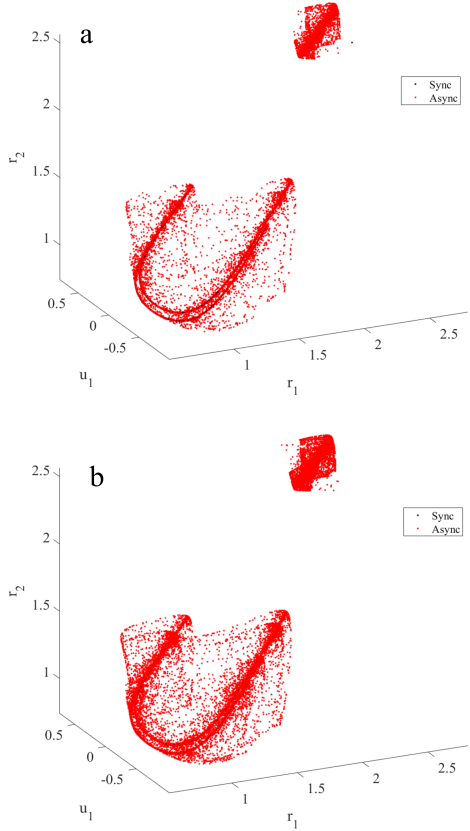}
}
\caption{Projections of Poincar\'e maps of attractors on the $(r_1, u_1, r_2)$ space at (a) $d/R_{10} = 12.30$ ($N_{async}/N = 0.636$, (b) $d/R_{10} = 12.36$, $N_{async}/N = 0.90$, the asynchronous component is dominating, the synchronous points can be barely seen behind the scattered asynchronous ones}
\label{fig6}
\end{figure}

In Fig.-s \ref{fig2}a-\ref{fig6}b we present projections of the Poincar\'e maps of the attractors at given parameter values on $(r_1, u_1, r_2)$ space. We use the Poincar\'e sections instead of phase curves, because high orbits density for the chaotic attractors makes it very hard to interpret the results. The corresponding values of Lyapunov exponents are presented in Tables \ref{tbl0}, \ref{tbl1}. Fig. \ref{fig2}a is the only one where we present the phase curve along with the Poincar\'e section, because it shows a synchronous limit cycle at $d/R_{10} = 7.6$. If one increases the parameter $d/R_{10}$, this limit cycle undergoes the cascade of period-doubling bifurcations and becomes a synchronous chaotic attractor. Poincar\'e section of this attractor  at $d/R_{10} = 12.01$ is presented in Fig. \ref{fig2}b. One can notice that all the points lie on the plane $r_1 = r_2$, which is the projection of the synchronization plane onto the three-dimensional space $(r_1, u_1, r_2)$.
Just as expected from a chaotic attractor embedded in the three-dimensional synchronization manifold, the largest Lyapunov exponent of this attractor is positive and the second one is negative, see Table \ref{tbl0}. Taking into account negative values of the corresponding largest transversal Lyapunov exponents for these synchronous attractors, we conclude that they are transversally stable on average and represent strongly synchronous states.

\begin{center}
\begin{table}[ht]
\caption{Lyapunov exponents and $N_{async}/N$ ratio for certain values of parameters.}
\begin{center}
\begin{tabular}{c|c|c|c|c|c|c|c|c|c}%{{1cm}|{1cm}|{1cm}|{1cm}|{1cm}|{1cm}|{1cm}|{1cm}|{1cm}|{1cm}|{1cm}|{1cm}}
Fig. & $d/R_{10}$ & $N_{async}/N$ & $\lambda_1$ & $\lambda_2$ & $\lambda^{sync}_1$ & $\lambda^{sync}_2$   & $\lambda^{async}_1$ & $\lambda^{async}_2$ & $\lambda_{tr}$ \\
\hline \\
\ref{fig2}a & $7.6$ & 0 & $-0.0114$ & $-0.0855$ & $-0.0114$ & $-0.0855$  & -- & -- & $-0.0855$ \\
\ref{fig2}b & $12.01$ & 0 & $0.0915,$ & $-0.0075$ & $0.0915$ & $-0.0075$ & -- & -- &  $-0.0080$ \\
\ref{fig3}a & $12.05$ & $ 0.716$ & $0.0503$ & $-0.0075$ & $-0.0559$ & $-0.1015$ & $0.0924$ & $0.0298$ & $0.0031$  \\
\ref{fig3}b & $12.12$ & $ 0.206$ & $0.0903$ & $ -0.0015$ & $0.0582$ & $-0.0326$ & $0.2143$ & $0.1189$ & $-0.0014$  \\
\end{tabular}
%\label{tbl1}
\end{center}
\label{tbl0}
\end{table}
\end{center}

If we further increase the value of $d/R_{10}$, we observe the start of the bubbling transition sequence (see the initial increase of $N_{async}/N$ and $\lambda_{tr}$ in Fig.-s \ref{fig1c}a,b inside $12.02 \lesssim d/R_{10} \lesssim 12.08$ interval). The left border of this quick growth marks the starting point at which subcritical pitchfork bifurcations start happening and the end of the existence of the synchronous attractor. We demonstrate in Fig. \ref{fig3}a an attractor that can be obtained at $d/R_{10} = 12.05$ and represents a typical attractor in this region of values of the control parameter. The whole orbit is characterized by one positive Lyapunov exponent, i.e. behavior on average is chaotic. However, the asynchronous component has two positive Lyapunov exponents implying hyperchaotic dynamics in this phase, while in the synchronous component all exponents are negative (see Table \ref{tbl0}). Notice also the $\lambda_{tr} > 0$ for this attractor, which means that it is transversally unstable on average and, consequently, corresponds to a weakly asynchronous state. While there still exist areas inside the synchronization manifold, where the Lyapunov sums are growing, the dynamics in the synchronous component on average become non-chaotic. This probably happens because the trajectory inside the synchronization manifold does not pass through the regions where the one-dimensional divergence of the orbits used to be the strongest in the synchronous attractor.

One can see that the transition from the synchronous state to weakly asynchronous one is not monotonous: after initial sharp increase of the asynchronous component, the number of asynchronous points on the attractor sharply drops almost to zero, and stays low for some time before beginning to grow again (see Fig. \ref{fig1c}a). In Fig. \ref{fig3}b we show an attractor at $d/R_{10} = 12.12$ that corresponds to the interval in $d/R_{10}$ for which the presence of the asynchronous phase is low, but still visible. In this case there are two positive Lyapunov exponents in the asynchronous phase and one positive Lyapunov exponent in the synchronous phase. On average, across both phases one exponent is positive and the others are negative. However, it is worth noting that for this attractor $N_{async}/N = 0.205$, which means that it belongs to the local spike in the area of low $N_{async}/N$ in Fig. \ref{fig1c}a. Negative value of $\lambda_{tr}$ for this attractor implies that it is weakly synchronous. Another attractor from the interval of low $N_{async}/N$, that is closer to the local minimum of asynchronous points, is presented in Fig. \ref{fig4}a and corresponds to $d/R_{10} = 12.18$. In this case $N_{async}/N = 0.037$ and only a few points belong the asynchronous phase. Thus, we do not present the values of Lyapunov exponents for the asynchronous component, because there are not enough points to make statistically correct conclusions. The values of Lyapunov exponents averaged across both phases and for synchronous phase are given in Table \ref{tbl1}. The behavior of an orbit is mostly determined by the synchronous phase and is chaotic on average. This attractor is also transversally stable on average.

\begin{center}
\begin{table}[ht]
\caption{Lyapunov exponents and $N_{async}/N$ ratio for certain values of parameters.}
\begin{center}
\begin{tabular}{c|c|c|c|c|c|c|c|c|c}%{{1cm}|{1cm}|{1cm}|{1cm}|{1cm}|{1cm}|{1cm}|{1cm}|{1cm}|{1cm}|{1cm}|{1cm}}
Fig. & $d/R_{10}$ & $N_{async}/N$ & $\lambda_1$ & $\lambda_2$ & $\lambda^{sync}_1$ & $\lambda^{sync}_2$   & $\lambda^{async}_1$ & $\lambda^{async}_2$ & $\lambda_{tr}$ \\
\hline \\
\ref{fig4}a & $12.18$  & $0.037$ & $0.0953$ & $-0.0032$ & $0.0836$ & $-0.0145$ & -- & -- & $-0.0027$ \\%$0.4042$ & $0.2949$ \\
\ref{fig4}b & $12.20$  & $0.150$ & $0.0952$ & $-0.0014$ & $0.0577$ & $-0.0385$ & $0.3083$ & $0.2093$ & $0.0001$ \\
%\ref{fig5}a & $12.24$ & $0.390$ & $0.0938$ & $-0.0002$ & $0.0449$ & $-0.0494$ & $0.1700$ & $0.0767$ \\
\ref{fig5}a & $12.27$  & $0.338$ & $0.0940$ & $-0.0006$ & $0.0528$ & $-0.0410$ & $0.1747$ & $0.0786$ & $0.0010$ \\
\ref{fig5}b & $12.275$ & $0.4389$ & $0.0936$ & $0.0004$ & $0.0294$ & $-0.0626$ & $0.1758$ & $0.0810$ & $0.0034$ \\
\ref{fig6}a & $12.30$  & $0.636$ & $0.0928$ & $0.0014$ & $-0.0101$ & $-0.0971$ & $0.1518$ & $0.0578$ & $0.0049$ \\
\ref{fig6}b & $12.36$  & $0.900$ & $0.0908$ & $0.0037$ & $-0.1033$ & $-0.1813$ & $0.1124$ & $0.0242$ & $0.0112$ \\
\end{tabular}
%\label{tbl1}
\end{center}
\label{tbl1}
\end{table}
\end{center}

In Fig.-s \ref{fig4}a--\ref{fig6}b we show a sequence of attractors with steadily growing amount of points belonging to the asynchronous phase. The values of Lyapunov exponents corresponding to each attractor are presented in Table \ref{tbl1}. Combined with the graphs from Fig.-s \ref{fig1a}a-\ref{fig1c}b, the general trend can be seen. In Fig. \ref{fig4}a the asynchronous component is barely seen, because there are only a few asynchronous points and the vast majority of them are located very close to the synchronization manifold. Thus, asynchronous points can be hardly visually distinguished from synchronous points. In Fig. \ref{fig4}b the asynchronous component becomes clearly noticeable, but still the asynchronous points are mostly concentrated in a narrow band that is close to the synchronization manifold. In these region of $d/R_{0}$ the values of $\lambda_{tr}$ oscillate close to zero, with a steady trend towards positive direction. This means that the synchronous attractor is on the verge of being transversally unstable.

Increasing $d/R_{10}$ further, we observe the next step of the proposed scenario. More and more transversally unstable saddle orbits of $(2, 2)$-type start appearing inside the synchronization manifold, which contributes into the formation of transversally unstable areas on the synchronization manifold. This can be observed in Fig.-s \ref{fig5}a,b. The number of asynchronous points in Fig. \ref{fig5}a is only two times larger than the number of those presented in Fig. \ref{fig4}b, however, they spread much further away from the synchronization manifold, which makes them significantly more apparent. The largest transversal Lyapunov exponent is positive for the last two attractors characterizing them as weakly asynchronous and keeps growing further with $d/R_{10}$.

In Fig. \ref{fig5}b the asynchronous phase is presented by even more far-off array of points. Not only they reach further away from the synchronization manifold, but also their density closer to the synchronization manifold increases significantly. At this point, the synchronous and asynchronous phases are roughly balanced in terms of time a trajectory spends in them, and by their impact on the overall dynamics on the attractor. The character of motion is very close to becoming hyperchaotic.

In the right side of the studied interval in Fig. \ref{fig1c}a, one can see that the attractor becomes almost asynchronous, i.e. $N_{async}/N$ becomes significantly larger than $N_{sync}/N$ and  the trajectory rarely comes back to the synchronization manifold, spending most of the time outside of it. Thus, after a certain point, the dynamics in the asynchronous phase almost completely determines the character of motion in the whole attracting area. Since the regimes of dynamics in the asynchronous phase are always hyperchaotic, the motion averaged across the whole attractor becomes hyperchaotic as well. This step of the proposed scenario corresponds to the last stage before the blowout bifurcation and corresponds to the scheme in Fig. \ref{fig0}c. Such attractors are presented in Fig.-s \ref{fig6}a,b. They correspond to $d/R_{10} = 12.30$ and $d/R_{10} = 12.36$ respectively, the values of the largest Lyapunov exponents and the largest transversal Lyapunov exponent can be found in Table \ref{tbl1}. From Fig.-s \ref{fig6}a,b we see that asynchronous points spread out significantly further away from the synchronization plane than before and the synchronous points are barely visible behind all the asynchronous ones. At $d/R_{10} = 12.30$, $N_{async}/N$ is equal to $0.636$ (see Fig. \ref{fig6}a), which means that there is still significant presence of the synchronous component, but nonetheless the motion is already hyperchaotic. From Fig. \ref{fig6}b one can observe that although visually asynchronous points do not extend further away for the synchronization manifold, in comparison to the asynchronous points in Fig. \ref{fig6}a, their density has become much higher. However, the trajectory still occasionally returns to the synchronization manifold after orbiting outside of it for a long time. This means that we do not observe the blowout bifurcation in our scenario. Since the largest transversal Lyapunov exponents are positive, these attractors correspond to weakly asynchronous states. Let us also remark that in the right side of the interval in Fig. \ref{fig1a}b the two largest Lyapunov exponents of the synchronous component start decreasing and both $\lambda_1^{sync}$ and $\lambda_2^{sync}$ become negative for attractors in Fig.-s \ref{fig6}a,b. This happens because the majority of the $(3, 1)$ orbits previously forming the skeleton of the synchronous attractor have become transversally unstable $(2, 2)$ orbits (making most parts of the synchronization manifold transversally unstable). Thus, a lot of these unstable orbits now contribute into the ejection of a trajectory outside of the synchronization manifold or prevent a trajectory from coming back into it, instead of contributing into the one-dimensional instability inside the synchronization manifold. In Fig. \ref{fig1c}a it is marked by the last growth of $N_{async}/N$ (with simultaneous drop of $N_{sync}$), after which $N_{async}/N \sim 0.9$ making $N_{sync}$ very small.

One can also notice the drop of the values of Lyapunov exponents at the very right end of the studied interval in Fig. \ref{fig1a}a. This corresponds to the crisis of the hyperchaotic attractor and a leap to an asynchronous limit cycle of period 4.

%\begin{widetext}
\begin{figure}[!ht]
\begin{center}
\includegraphics[width=0.5\textwidth]{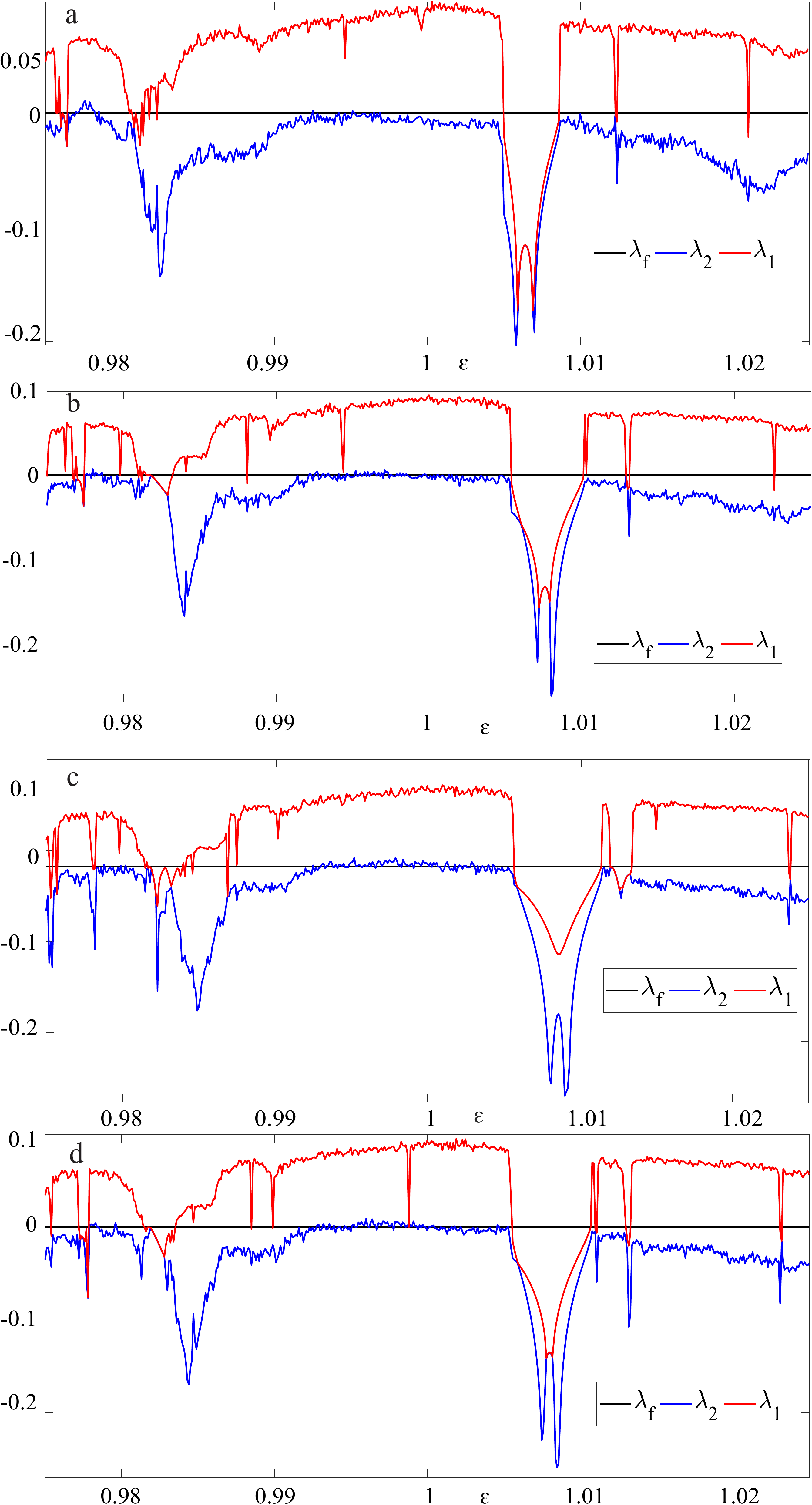}
\end{center}
\caption{The dependence of Lyapunov spectra on $\varepsilon$ for continuation of (a) the completely synchronous attractor attractor at $d/R_{10} = 12.01$, (b) the attractor the in middle of the bubbling transition process at $d/R_{10} = 12.20$, (c) the attractor in the middle of the bubbling transition at $d/R_{10} = 12.27$, (d)  the hyperchaotic attractor in the end of the bubbling transition at  $d/R_{10} = 12.36$.}
\label{f:fig7}
\end{figure}
%\end{widetext}

We would also like to note that the attractor we are studying here is relatively stable to symmetry breaking in model \eqref{eq:eq1}, corresponding to the transition from two identical bubbles to a pair of slightly different. Let us introduce a small parameter that characterizes this transition as follows\cite{Garashchuk2020} $R_{20} = \varepsilon R_{10}$, $|\epsilon-1| \ll 1$, $\varepsilon>0$. If $\varepsilon \neq 1$, equations \eqref{eq:eq1} are no longer symmetrical with respect to $R_1 \longleftrightarrow R_2$ transformation.

Now we study the influence of symmetry breaking on several attractors representing typical stages of the bubbling transition process. We numerically continue these attractors with respect to $\varepsilon$ and calculate the corresponding Lyapunov spectra. We present the results of this continuations for four attractors along the way of the bubbling transition scenario: the synchronous chaotic attractor from Fig. \ref{fig2}b, $d/R_{10} = 12.01$, (see Fig. \ref{f:fig7} a); the one with low presence of the asynchronous component from Fig. \ref{fig4}b, $d/R_{10} = 12.20$ (see Fig. \ref{f:fig7}b); the attractor with moderate presence of the asynchronous component from Fig. \ref{fig5}a, $d/R_{10} = 12.27$ (see Fig. \ref{f:fig7}c); the hyperchaotic attractor for which the asynchronous component is dominant from Fig. \ref{fig6}b, $d/R_{10} = 12.36$ (see Fig. \ref{f:fig7}d). The results for all these attractors are quite similar, except that dynamics is hyperchaotic in some neighborhood of $\varepsilon = 1$ on the last two graphs.

In \cite{Garashchuk2020} attractors were considered unstable with respect to symmetry breaking if they disappear through a crisis or underwent other sharp qualitative changes of dynamics almost instantly after the symmetry breaking is introduced (i.e. $| \varepsilon - 1|  \ll 0.01$). This corresponds to deviations of the equilibrium radii from the symmetrical state approximately on the order of $0.1\%$ or less. On the other hand, if an attractor is stable with respect to perturbation in $\varepsilon$ on the order of $1\%$ or more  (i.e. $\varepsilon \sim 1 \pm 0.01$), it was considered stable to symmetry breaking because this is a sensible margin of error from a physical point of view. Here we follow the same approach. Graphs in Fig.-s \ref{f:fig7}a-d are very similar, and the sharp qualitative change in type of behavior can be observed at the values of $\varepsilon$ slightly smaller then $1.01$, which corresponds to deviations from symmetrical case on the order of $0.7-0.8\%$. These regions of $\varepsilon$, though relatively small, can be considered as areas of stability with respect to symmetry breaking in comparison with the cases when attractors disappear with arbitrary small perturbations in $\varepsilon$. Therefore, we consider that all the discussed above attractors are stable with respect to symmetry breaking. As a consequence, dynamical regimes emerging in the entire bubbling transition scenario might be observed in a system of two slightly nonidentical contrast agents and the bubbling transition process may be observed experimentally.

Finally, we would like to note that the bubbling transition provides a new bifurcation scenario for the appearance of hyperchaos in system \eqref{eq:eq1}. This scenario complements the one that is based on the appearance of a homoclinic chaotic attractor containing a saddle-focus periodic orbit with its two-dimensional unstable manifold\cite{Garashchuk2019} (see, also\cite{Stankevich2018,Stankevich2019} for other applications of the latter scenario).

\section{Conclusion}
\label{sec:end}

In this work we have studied the process of the destruction of a synchronous chaotic attractor in a model of oscillations of two interacting identical ultrasound contrast agents. We have phenomenologically described the main steps and driving mechanisms of the corresponding scenario. We have numerically shown that the implementation of such bifurcation scenario takes place in the studied model.

We have found that trajectories can be ejected from the synchronization manifold, which confirms the existence of transversal instabilities inside it. Moreover, we have demonstrated that the asynchronous phase is characterised by two positive Lyapunov exponents, which implies the presence of $(2, 2)$-type saddle orbits in the saddle set outside of the synchronization manifold. Thus, this asynchronous saddle set is hyperchaotic. We have provided Poincar\'e maps of individual attractors that highlight certain steps of the scenario and support the general picture shown in Fig.-s \ref{fig1a}a-\ref{fig1c}b.

Implementation of the bubbling transition scenario discussed here is similar to the one presented in work\cite{Saha17}. However, in\cite{Saha17} the asynchronous saddle set was simply chaotic. In our case this set is hyperchaotic and, as a result, motion in the asynchronous phase is also hyperchaotic. Consequently, proposed scenario can be considered as another route to hyperchaos in systems of coupled oscillators.

We have also studied the stability of the proposed scenario with respect to the symmetry breaking perturbation. We have numerically shown that several attractors at the  main steps of the scenario are stable with respect to symmetry breaking  and their behaviour after symmetry breaking is quite similar. This means that the entire scenario could be observed at values of $\varepsilon$ close but not equal to 1, that correspond to the case of two bubbles with slightly different equilibrium radii. This suggestion is a little counterintuitive, since for $\varepsilon \neq 1$ the symmetry of system \eqref{eq:eq1} no longer exists together with the synchronous attractors and the $Fix(S)$ manifold, that play an important role in the scenario described in Section \ref{sec:scenario}. Nevertheless, it seems that small shifts and deformations of the orbits in Fig. \ref{fig0} do not eliminate the possibility of implementation of the scenario. Therefore, we believe that the dynamical regimes and scenarios discussed here may be observed in physically realistic systems.

We would also like to note that hyperchaotic oscillations of contrast agents may be beneficial for various applications, for example, for applications related to targeted drug delivery, when a quick rupture of the bubbles is desirable\cite{Hoff,Klibanov2006,Carroll2013}. Moreorver, hyperchaotic attractors are more robust to various perturbations including symmetry breaking. Therefore, we believe that finding new routes to hyperchaos may be important for certain applications of the contrast agents.

\section*{Acknowledgments}
Authors are grateful to Alexey Kazakov and anonymous referees for useful discussions and remarks that helped to significantly improve the manuscript. This work (except for Section 3) was supported by Russian Science Foundation grant no. 19-71-10048. The results in Section 3 were supported by Laboratory of Dynamical Systems and Applications NRU HSE, of the Ministry of science and higher education of the RF grant ag. N. 075-15-2019-1931

\end{document}